\documentclass{iaus}
\usepackage{graphicx}

\title[ISM in ULIRGs] 
{The irradiated ISM of ULIRGs}

\author[Spaans et al.]   
{M. Spaans$^1$, R. Meijerink$^2$, F.P. Israel$^3$, A.F. Loenen$^{1,4}$, W.A. Baan$^4$}

\affiliation{$^1$Kapteyn Astronomical Institute, P.O. Box 800, 9700 AV Groningen, The Netherlands \break email: spaans@astro.rug.nl\\[\affilskip]
$^2$Astronomy Department, University of   California, Berkeley, CA 94720, United States\\[\affilskip]
$^3$Leiden Observatory, P.O. Box 9513, 2300 RA Leiden, The Netherlands\\[\affilskip]
$^4$ASTRON, P.O. Box 2, 7990 AA Dwingeloo, The Netherlands
}

\pubyear{2007}
\volume{242}  
\pagerange{--}
\date{?? and in revised form ??}
 \jname{Astrophysical Masers and their
Environments} \editors{J.M. Chapman \& W.A. Baan}
\begin{document}

\maketitle

\begin{abstract}
The nuclei of ULIRGs harbor massive young stars, an accreting
central black hole, or both. Results are presented for molecular
gas that is exposed to X-rays (1-100 keV, XDRs) and
far-ultraviolet radiation (6-13.6 eV, PDRs). Attention is paid to
species like HCO$^+$, HCN, HNC, OH, H$_2$O and CO. Line ratios of
HCN/HCO$^+$ and HNC/HCN discriminate between PDRs and XDRs. Very
high $J$ ($>10$) CO lines, observable with HIFI/Herschel,
discriminate very well between XDRs and PDRs. In XDRs, it is easy
to produce large abundances of warm ($T>100$ K) H$_2$O and OH. In
PDRs, only OH is produced similarly well.

\keywords{ISM: molecules -- galaxies: evolution}
\end{abstract}

\firstsection 
\section{Introduction}

The power that emanates from (ultra-)luminous infrared galaxies is
believed to stem from active star formation and/or an accretion
disk around a central super-massive black hole (e.g.\ Sanders \&
Mirabel 1996). The unambiguous identification of the central
energy source, or the relative contributions from stars and an
active galactic nucleus (AGN), remains a major challenge in the
study of active galaxy centers (Aalto et al.\ 2007). Molecular
lines are ideal in this to penetrate deep into the large column
density regions of active galaxies (Israel 2005; Baan \&
Kl\"ockner 2005). The interested reader is referred to Meijerink
\& Spaans (2005) and Meijerink, Spaans \& Israel (2006, 2007) for
details on Photon-Dominated Region (PDR) and X-ray Dominated
Region (XDR) physics. PDRs refer to the presence of a starburst
that produces photons with energies of 6-13.6 eV, while XDRs
indicate an accreting black hole with photon energies of 1-100 keV
(Maloney et al.\ 1996; Lepp \& Dalgarno 1996). The most important
thing to realize is that a 1 keV photon penetrates a hydrogen
column of about $10^{22}$ cm$^{-2}$, while a UV photon (10 eV) is
absorbed by dust after about 1 mag of visual extinction. This is a
consequence of the fact that X-ray absorption cross sections scale
roughly like energy$^{-3}$, allowing deep penetration of X-rays
into interstellar clouds.

\section{Results}

All presented models are plane-parallel slabs and are
parameterized by a constant density and an impinging UV (in
multiples of $G_0=1.6\times 10^{-3}$ erg s$^{-1}$ cm$^{-2}$) or
X-ray (energy$^{-0.9}$) radiation field $F_X$ in erg s$^{-1}$
cm$^{-2}$. Note that a flux of $F_X=100$ erg s$^{-1}$ cm$^{-2}$
corresponds to a $10^{44}$ erg s$^{-1}$ Seyfert nucleus at 100 pc
from an interstellar cloud.

\begin{figure}
\includegraphics[height=90mm,width=120mm]{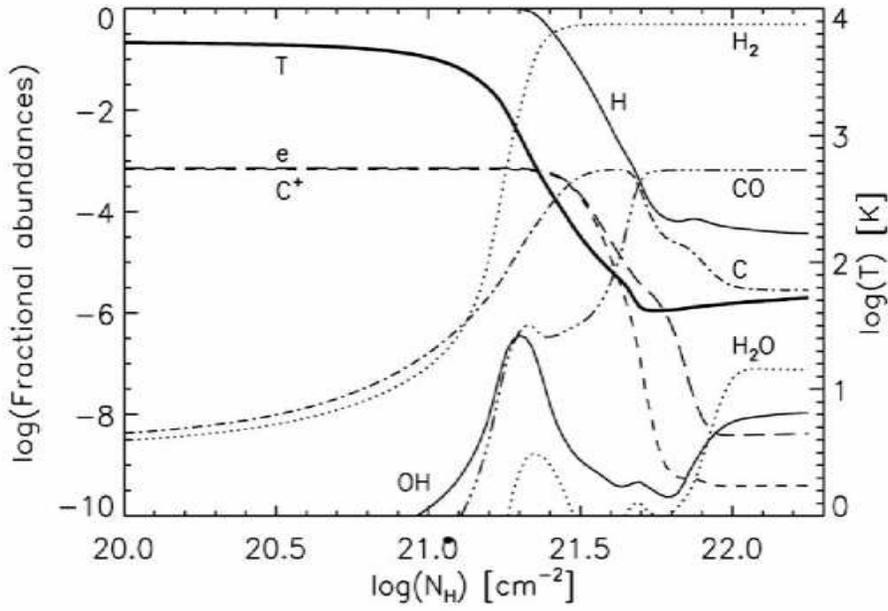}
  \caption{Depth dependence of a few important chemical species in the
  PDR, $n=10^5$ cm$^{-3}$ and $10^{3.5}G_0$. Note the clear
  stratification.}\label{F1}
\end{figure}

\begin{figure}
 \includegraphics[height=90mm,width=120mm]{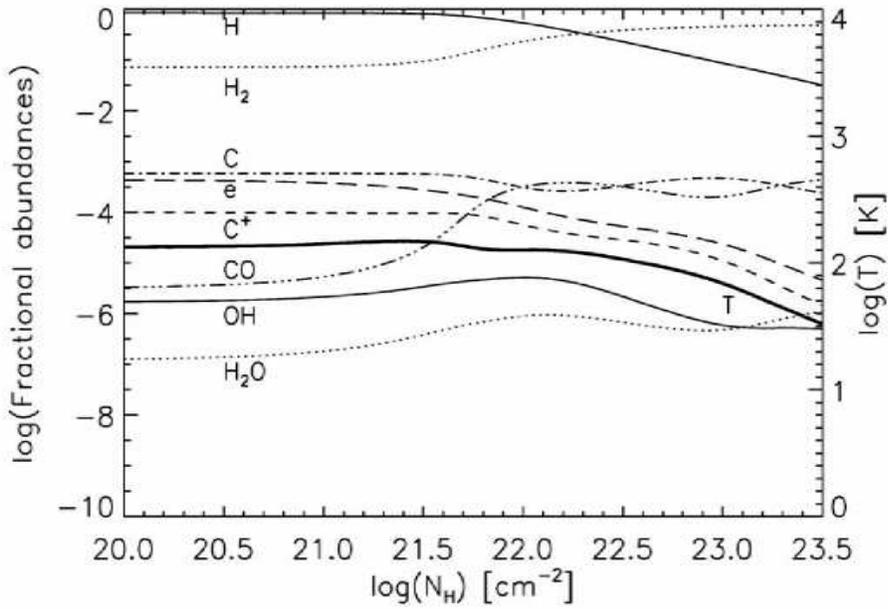}
  \caption{Depth dependence of a few important chemical species in
  the XDR, $n=10^5$ cm$^{-3}$ and $F_X=5$ erg s$^{-1}$ cm$^{-2}$.
  Note the different column density scale and warm water.}\label{F2}
\end{figure}


It is clear from figures 1, 2, 3 and 4 that OH is easily formed in
PDRs and XDRs, while warm water is present mostly in XDRs. In
PDRs, OH is the dissociation product of water, while it is formed
indirectly through charge transfer of H$^+$ and O, rapid reactions
with H$_2$ to OH$^+$ and H$_2$O$^+$, followed by dissociative
recombination or directly through O+H$_2$$\rightarrow$ OH+H. The
latter reaction is driven efficiently above 200 K. Additional
reactions of H$_2$O$^+$ with molecular hydrogen lead to H$_3$O$^+$
which dissociatively recombines to, among others, water or OH.
Note that an elevated cosmic ray ionization rate also leads to
larger abundances of H$_3$O$^+$, but not to the same degree as in
XDRs (Meijerink et al.\ 2006). In both PDRs and XDRs,
vibrationally excited H$_2$ is also present through, respectively,
UV pumping and thermal collisions with electrons. This
significantly lowers the effective energy barrier of the O+H$_2$
reaction. In XDRs, an internal UV radiation field is created by
collisional excitation of H and H$_2$ that leads to Lyman $\alpha$
and Lyman-Werner photons through radiative decay. However, this UV
field is not strong enough to dissociate water as efficiently as
in PDRs. At the high temperatures reached in XDRs, because
photo-ionization heating is more efficient than photo-electric
heating by dust grains, the endo-ergic reaction OH+H$_2$ also
leads to water. Of course, shocks are similarly capable of
producing OH and H$_2$O.

\begin{figure}
 \includegraphics[height=90mm,width=120mm]{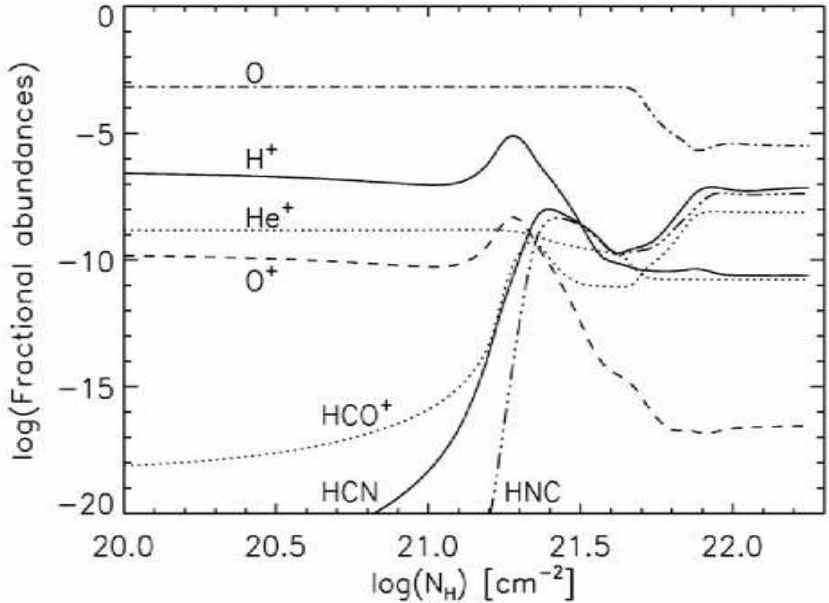}
  \caption{Depth dependence of a few important species in the PDR, same as
  figure 1.}\label{F3}
\end{figure}


Typically, the HCO$^+$ lines are stronger in XDRs than in PDRs by
a factor of at least three. This is a direct consequence of the
higher ionization degree in XDRs (Meijerink \& Spaans 2005),
leading to an enhanced HCO$^+$ formation rate. Depending on the
incident radiation field, HCN or HCO$^+$ is more abundant at the
PDR edge of the cloud.
 At sufficiently large column and densities, the
HCN(1-0)/HCO$^+$(1-0) ratio becomes larger than 1. In the XDR
models, HCO$^+$ is chemically less abundant than HCN for very
large $H_X/n$ (Meijerink \& Spaans 2005). However, for larger
columns HCO$^+$ always becomes more abundant than HCN (Fig.\ 10 in
Meijerink \& Spaans 2005).
 These effects are summarized in figure 2, 4 and 5, see also
Loenen et al.\ (this volume).

\begin{figure}
 \includegraphics[height=90mm,width=120mm]{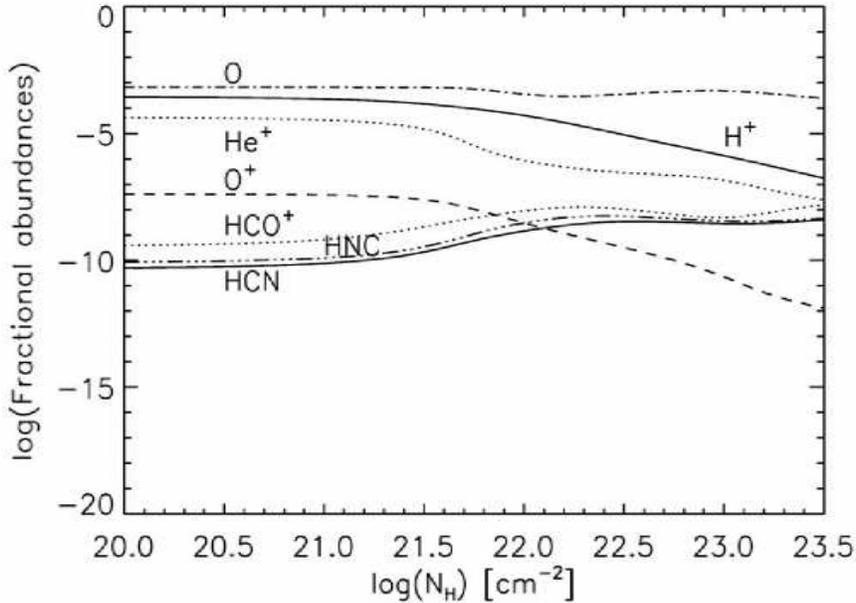}
  \caption{Depth dependence of a few important species in the XDR, same as
  figure 2. Note the enhancement of HNC with respect to HCN (factor
  2).}\label{F4}
\end{figure}


The critical densities of HCN and HNC are almost identical, so the
only differences in ratios can be due to differences in the
abundances. It turns out that the HNC/HCN column density ratio is
quite close to that of HNC(1-0)/HCN(1-0) line intensity ratio. In
PDRs, HCN is more abundant in the radical region, but deeper into
the cloud the ratio becomes about one. In XDRs, HCN is more
abundant in the highly ionized part of the cloud. However, HNC is
equally or even more abundant than HCN deep into the cloud. All
this can be seen in figures 2 and 4.

In XDRs at high densities CO is present all throughout the cloud,
even when $F_X/n$ is large. This is a direct consequence of the
fact that X-rays do not lead to strong dissociation of CO and thus
C$^+$, C and CO typically co-exist in an XDR at elevated ($>100$
K) temperatures. Such warm CO gas produces emission originating
from high rotational transitions. Contrary to XDRs, most CO in
PDRs is produced after the H/H$_2$ transition and has on average
much lower temperatures ($T\sim20-50$~K). Future missions such as
Herschel/HIFI will be able to distinguish between PDRs and XDRs by
observing high rotational transitions such as CO(16-15), see
figure 6.

\begin{figure}
 \includegraphics[height=135mm,angle=-90]{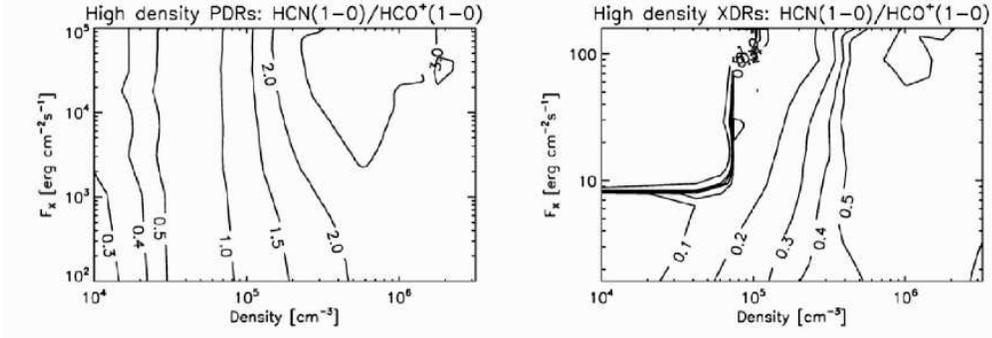}
  \caption{The HCN(1-0)/HCO$^+$(1-0) line intensity ratio for a grid of PDR
  (left) and XDR (right) models.}\label{F6}
\end{figure}


\begin{figure}
 \includegraphics[height=135mm,angle=-90]{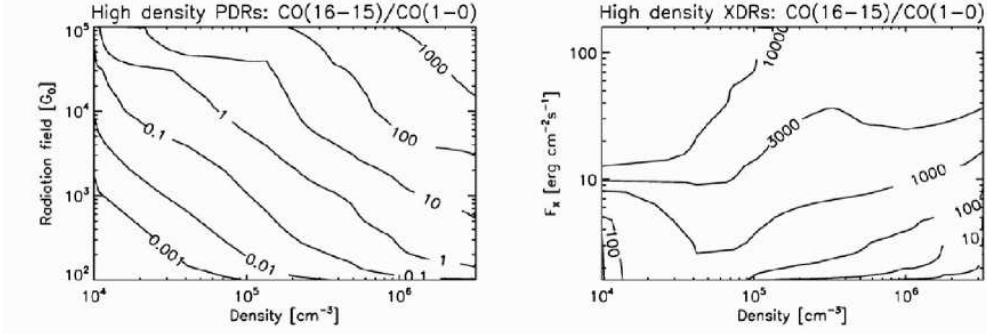}
  \caption{The CO(16-15)/CO(1-0) line intensity ratio for a grid of PDR (left)
  and XDR (right) models.}\label{F8}
\end{figure}

\section{Future work}

In the future, observatories like Herschel and ALMA will
revolutionize our understanding of the nuclear activity in (maser)
galaxies by resolving the central regions of these systems
spatially. In this light, it is important to stress that starburst
activity typically occupies a larger fraction of an active galaxy
than accretion onto a central black hole. Consequently, single
dish observations are likely to be dominated by a PDR signal even
when an XDR is present. This is particularly true for the very
high $J$ CO lines and emphasizes the need for high spatial
resolution. ALMA can detect and resolve very high $J$ CO emission
from ULIRGs at redshifts beyond eight. Additional molecules that
allow one to probe an accreting black buried inside ULIRGs, i.e.\
that are direct tracers of X-ray irradiation, are CO$^+$, CH$^+$
and H$_3^+$.

\begin{acknowledgments}
We are grateful to Dieter Poelman and Juan Pablo P\'erez-Beaupuits
for discussions.
\end{acknowledgments}




\end{document}